# Analytic Design of Flat-Wire Inductors for High-Current and Compact DC-DC Converters

Sajjad Mohammadi
*Electrical Engineering and Computer Science*
*MIT*
Cambridge, USA
sajjadm@mit.edu

James L. Kirtley
*Electrical Engineering and Computer Science*
*MIT*
Cambridge, USA
kirtley@mit.edu

Alireza Namadmalan
*Power Electronics & Magnetics Lab*
*Bourns Electronics Ireland (BEI)*
Cork, Ireland
alireza.namadmalan@bourns.com

*Abstract*—This paper presents analytic study and design considerations of flat wire inductors with distributed gaps for high-power and compact DC-DC Converters. The focus is eddy-current loss components within the conductors due to fringing and leakage fluxes. A magnetic equivalent circuit (MEC) is proposed in which eddy currents are modeled by MMFs opposing the primary flux as well as frequency dependent reluctances, which finally leads to a frequency dependent inductance describing the behavior of the inductor at high frequencies. Three formulations for DC resistance depending on the required accuracy are developed. Calculations of the AC resistance based on vector potential obtained from FEM are provided. To provide an insight into the optimized design of such inductors, components of the magnetic flux and induced eddy currents along with sensitivity of the main inductor quantities such as DCR, ESR, loss components and inductance values to the design parameters are investigated. Finally, an inductor is prototyped and experimentally tested to verify the design.

## I. Introduction

Recent advances in high-frequency and wide-band-gap semiconductors, such as GaN and SiC switches, have prioritized the miniaturization of power converters [1], where magnetic components, especially high-frequency inductors are usually the limiting components in terms of size and losses [2]-[5]. Typically, there are applications where current of high-power inductors include both high-frequency and low-frequency (or DC) components. These applications include DC-DC inductors, power factor correction (PFC) inductors, output filter inductors, and chokes [4]-[7]. The main challenge in these designs lies in selecting and designing the winding to manage both low and high-frequency conduction losses effectively [8].

Litz wire is a common solution to this challenge, yet its poor filling factor of about 50%, or sometimes less, limits its effectiveness for high power density applications. Compared to solid copper wires, litz wires leads to bulkier inductors with higher direct current resistance (DCR), negatively impacting low-frequency or DC performance. However, poorly designed solid wires can lead to increased equivalent series resistance (ESR) at high frequencies [8]. Various solutions using solid copper wires aim to optimize the balance between DCR and ESR at high frequencies. For example, a design using single-layer winding of round wires can reduce ESR for high-frequency applications [9]. Although it works well for toroidal shapes, it presents high ESR for multi-layer designs [10].

Solid wires with spiral shape, although promising, are mainly suitable for PCB inductors [2]. Solid flat wires, although provide a better thermal conductivity and DCR, require careful design optimization [7]. Among solid wire options, flat spiral and helix shapes offer superior performance and offer more practical manufacturing on printed circuit boards (PCBs), helping to reduce the size of the magnetic components. Spiral shapes, though effective, have proximity effect issues and are more suitable for high-frequency planar transformers [2].

Beside superior benefits for solid flat wire inductors, still they require further research in terms of accurate modelling [7], and [10]. To do design optimization of such inductors, e.g. using heuristic methods presented in [12], accurate models are needed to derive analytical equations for AC and DC losses in the solid flat wire coils. The model should consider the frequency-dependent effects of eddy-currents, reluctance paths and fringing effect of the core's gaps, which has not been proposed for such inductors. As an example, [13]-[14] has presented such a model for eddy currents in the lamination and magnets of an actuator, where eddy current impacts are represented as frequency-dependent reluctance and inductance.

The main contribution of this paper is presenting an analytic study and design considerations of high-power inductors with helical flat wires and distributed gaps with a focus on eddy-current effects in the coil. A magnetic equivalent circuit (MEC) including magneto-motive forces or frequency-dependent reluctances is offered which explain the inductance reduction at higher frequencies. Equations for DC and AC resistances for three cases are developed. FEM is employed in the analyses. Flux and eddy current components as well as the impact of design parameters on the performance indices of the inductor are investigated. Finally, an inductor with PQ 40/40 ferrite core is prototyped and tested to experimentally verify the design.

## II. Inductor Topology

Fig. 1 shows an inductor including a PQ 40/40 core with a distributed gap and a flat-wire coil. As the fringing fluxes almost take a circular path with an effective radius of the gap length, as illustrated in Fig. 2, distributing a large gap into multiple small gaps leads to less penetration of the fringing flux into the window, resulting in less eddy current loss.

### III. COUPLED MAGNETIC-ELECTRIC CIRCUIT

Figs. 3 (a)-(b) show flux lines, flux density and field intensity within the inductor for a DC current of 5 A with no eddy-current effect. It is a two-dimensional FEM with a symmetry along ϕ axis. We have flux fringing at the gaps and a leakage through the window. Field intensity distribution shows that most of the stored energy is mainly within the air-gap, and then the fringing regions.

Accordingly, magnetic equivalent circuit shown in Fig. 4(a) can be developed in which $R_g$, $R_{ci}$, $R_f$, and $R_{lw}$ are gap reluctance, $i^{th}$ core reluctance, fringing reluctance and window leakage reluctance. As shown in Fig. 4(b), eddy currents induced in the conductors by fringing or window leakage fluxes can be modeled by magneto-motive forces $F_{ef}$ and $F_{ew}$, respectively, which generate fluxes $\phi_{ef}$ and $\phi_{ew}$ in the opposite direction of the original flux generated by the coil in the fringing region and the window, respectively. As represented in Fig. 4(c), it can be shown that these MMFs can be modeled by frequency dependent reluctances $R_{ef}(j\omega)$ and $R_{elw}(j\omega)$ in series with the primary zero-frequency reluctance. These frequency-dependent reluctances go up at higher frequencies which explains why inductance goes down by a reduction in the total flux linked by the coil at higher frequencies [13]-[14].

$$\varphi(j\omega) = \frac{Ni_L}{R_t(j\omega)} = \frac{Ni_L}{R_0 + \underbrace{R_0 Q(j\omega)}_{R_e(j\omega)}} = \frac{Ni_L}{R_0(1+Q(j\omega))} \quad (1)$$

where $R_t$ is the total reluctance, $R_0$ is the total reluctance at zero frequency and $R_e$ is the total reluctance added by eddy currents which is zero at zero frequency with Q as an auxiliary term:

$$\omega \to 0 \Rightarrow Q(j\omega) \to 0;\ R_e(j\omega) \to 0;\ R_t(j\omega) \to R_0;\ \varphi(j\omega) \to \varphi_0 \quad (2)$$

where $\phi_0$ is the primary flux at zero frequency:

$$\varphi_0 = \frac{Ni_L}{R_0} \quad (3)$$

The coupled electric-magnetic circuit for both zero-frequency case and the frequency-dependent case are shown in Fig. 5. The governing equations can be expressed as:

$$\begin{cases} V_L = R_L I_L + j\omega N\varphi \\ NI_L = R_t \varphi \end{cases} \Rightarrow \begin{bmatrix} R_L & j\omega N \\ -N & R_t \end{bmatrix} \begin{bmatrix} I_L \\ \varphi \end{bmatrix} = \begin{bmatrix} V_L \\ 0 \end{bmatrix} \quad (4)$$

There is a codependency between electrical and magnetic circuits. From the magnetic circuit, flux is returned to the electrical circuit, and from the electrical circuit, the current is returned to the magnetic circuit. By solving the above system of equation, the terminal impedance can be obtained as:

$$Z(j\omega) = \frac{V_L}{I_L} = R_L + j\omega L(j\omega);\ L(j\omega) = \frac{N^2}{R_t(j\omega)} = \frac{L_0}{1+Q(j\omega)} \quad (5)$$

It can be seen that the inductance $L(j\omega)$ is $L_0$ at zero frequency, and at higher frequencies, as Q and subsequently the reluctance goes up, the inductance goes down.

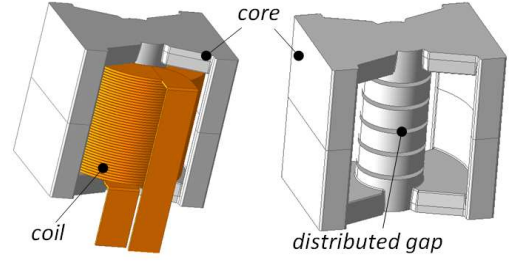
Fig. 1. Topology (top) and exploded view (bottom) of the proposed coupler.

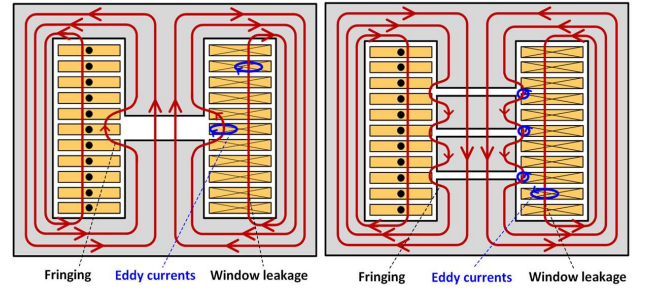
Fig. 2. Flux paths for inductors with single and distributed gaps.

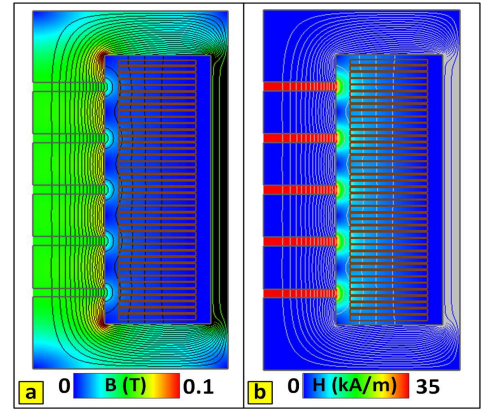
Fig. 3. Flux lines, flux density and field intensity with the core and window.

### IV. DC RESISTANCE AND LOSS

Three models for calculating DCR are represented here that can be employed depending on the inductor topology and the required accuracy.

#### A. Coil with Helical Shape

Fig. 6 shows the geometry of the coil in which $R_w$, $D_w$, $t_w$, $h_w$, $s$, $N$ are inner radius of the coil, radial length of the coil, thickness or z-height of the flat wire, total height of the coil, distance between turns, and number of turns, respectively. Moreover, $D_{righ}$ and $D_{left}$ are clearance between coil and core, shown in Fig. 6. A differential area of $dA = t_w dr$ for a differential radius of $dr$ is employed in the integration. Using the length of helix $l(r)$ with radius $r$, $N$, and a total height of $h_w = Nt_w + (N-1)s$, the conductance is obtained as:

$$G_{DC} = \frac{1}{R_{DC}} = \int_{r=r_w}^{r=r_w+D_w} \frac{\sigma t_w dr}{l(r)} \quad (6)$$

where the total length of the coil with N turns is as follows:

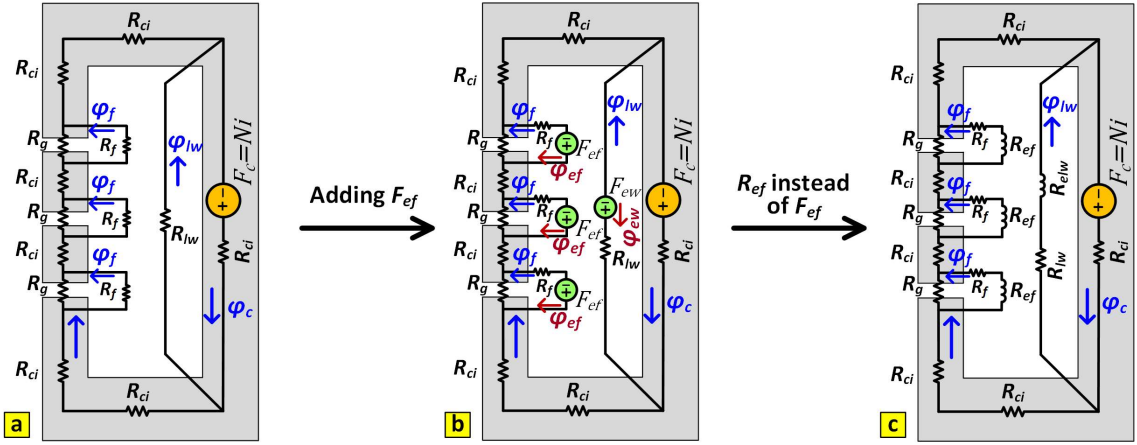

Fig. 4. MEC development: (a) eddy-currents ignored, (b) MMFs $F_e$ representing eddy-currents an opposing flux, and (c) frequency-dependent reluctances $R_e$.

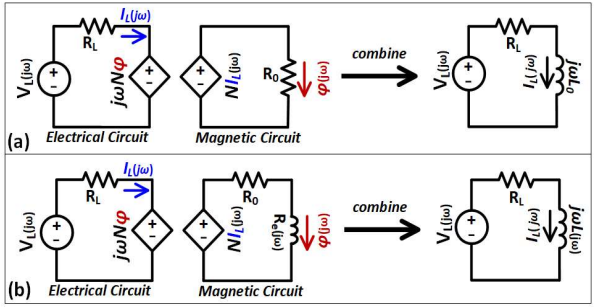

Fig. 5. Coupled electric-magnetic circuit: (a) low frequency case with no frequency dependency, and (b) frequency dependency due to eddy currents.

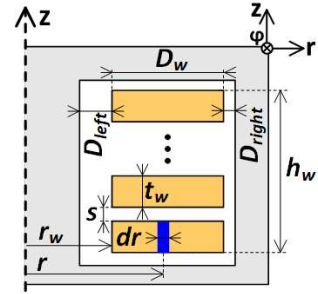

Fig. 6. Coil dimensions.

$$l(r) = 2\pi N \sqrt{r^2 + \left(\frac{h_w}{2\pi N}\right)^2} \quad (7)$$

Finally, DCR is obtained as:

$$R_{DC} = 2\pi N \left/ \left\{ \sigma t_w \ln\left(\frac{r_w + D_w + \sqrt{(r_w + D_w)^2 + \left(\frac{h_w}{2\pi N}\right)^2}}{r_w + \sqrt{r_w^2 + \left(\frac{h_w}{2\pi N}\right)^2}}\right) \right\} \right. \quad (8)$$

This is the general and the most accurate relationship for calculation of the DCR of this type of coil which is particularly useful for small number of turns where $h_w/2\pi N$ is not small compared to $r_w+D_w$.

### B. Planar Circular Approximation

For large number of turns, coil length can be approximated by discrete circular turns as follows. It can also be obtained from approximation of the previous equation if $h_w/2\pi N$ is small enough and negligible. The conductance can be derived as:

$$G_{DC} = \frac{1}{R_{DC}} = \int_{r=r_w}^{r=r_w+D_w} \frac{\sigma t_w dr}{l(r)} \quad (9)$$

where the total length of the coil with $N$ turns is as follows:

$$l(r) = 2\pi N r \quad (10)$$

Finally, DCR is obtained as:

$$R_{DC} = 2\pi N \left/ \left\{ \sigma t_w \ln\left(\frac{r_w + D_w}{r_w}\right) \right\} \right. \quad (11)$$

### C. Average Radius Approximation

An initial estimation of DCR can be obtained by cconsidering an average radius $R_{av}=r_w+D_w/2$ as follows:

$$G_{DC} = \frac{1}{R_{DC}} = \int_{r=r_w}^{r=r_w+D_w} \frac{\sigma t_w dr}{l(r)} \quad (12)$$

where the total length of the coil with $N$ turns is as follows:

$$l(r) = 2\pi N R_{av} \quad (13)$$

Finally, DCR is obtained as:

$$R_{DC} = \frac{2\pi N R_{av}}{\sigma t_w D_w} \quad (14)$$

## V. AC RESISTANCE AND CONDUCTION LOSSES

This section is devoted to the calculations of AC resistance of the inductor at high frequencies. According to Gausses' law $\nabla \cdot B = 0$, a magnetic vector potential $A$ can be defined as $B = \nabla \times A$. Employing the identity $\nabla \times \nabla \times A = \nabla(\nabla \cdot A) - \nabla^2 A$ and Coulomb's gauge condition $\nabla \cdot A = 0$ in Ampere's law yields:

$$\nabla \times H = J \xrightarrow{H=B/\mu_0} \nabla \times \left(\frac{\nabla \times A}{\mu_0}\right) = J \rightarrow \nabla^2 A = -\mu_0 J \quad (15)$$

In the two-dimensional domain with symmetry in the $\phi$ direction, the vector potential $A_\phi$ and the currents $J_\phi$ are in the

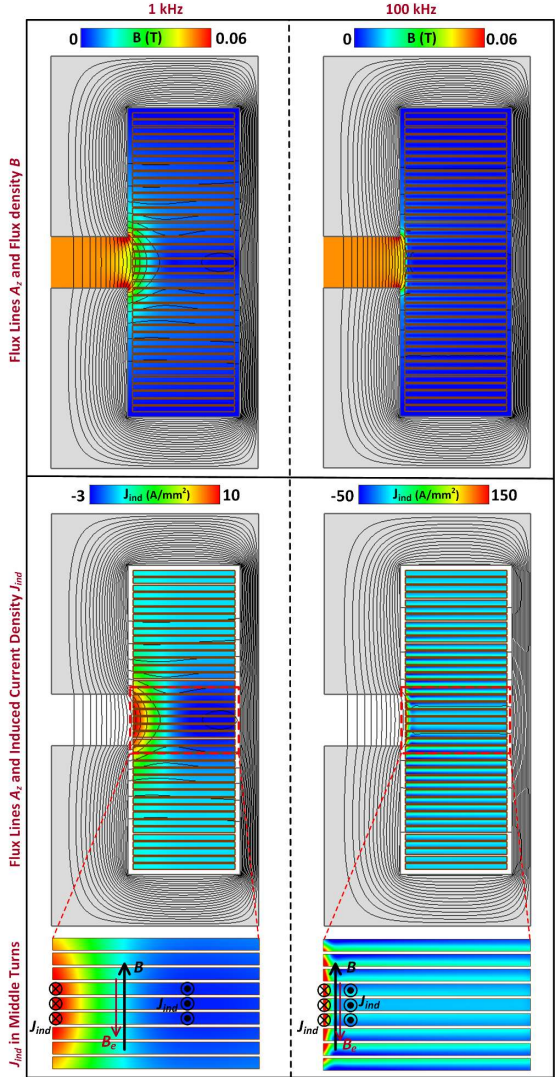

Fig. 7. Flux lines, flux density, and induced eddy currents in the conductors at frequencies of 1 kHz and 100 kHz for inductor with single large gap.

$\phi$ direction, while magnetic fields $B$ and $H$ are in the $rz$-plane. We have:

$$\nabla^2 A_\varphi(r,z,t) = -\mu_0 J_\varphi(r,z,t) \quad (16)$$

It is a 2-D problem with $\varphi$-directed current $J_\varphi$ that can be decomposed into the source $J_s(t)$ due to inductor terminal current, and also induced eddy-current density $J_{eddy}$. As follows:

$$J_\varphi(r,z,t) = J_s(t) + J_{eddy}(r,z,t) \quad (17)$$

The inductor terminal current $I_L$ can be obtained by integrating $J_s$ which is uniformly distributed over a conductor area as follows:

$$I_L(t) = \iint_{A_w} J_s(t)\, dr\, dz = t_w D_w J_s(t) \quad (18)$$

Using Faraday's law and Ohm's law $J_\varphi = \sigma E_\varphi$, induced eddy current density can be obtained in terms of $A_\varphi$ as follows:

$$\nabla \times E_\varphi(r,z,t) = -\frac{\partial B_{rz}(r,z,t)}{\partial t} \xrightarrow{B=\nabla \times A} E_\varphi(r,z,t) = -\frac{\partial A_\varphi(r,z,t)}{\partial t} \quad (19)$$

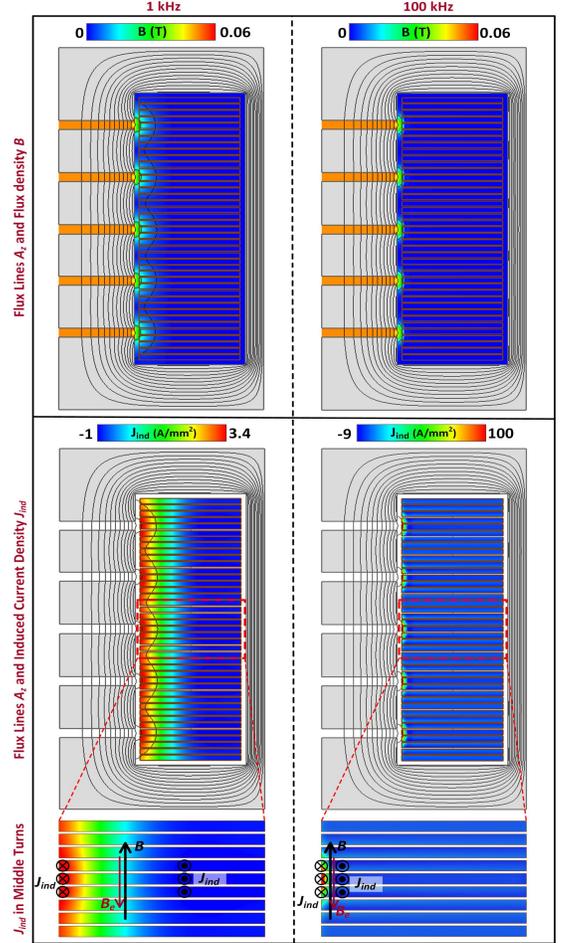

Fig. 8. Flux lines, flux density, and induced eddy currents in the conductors at frequencies of 1 kHz and 100 kHz for distributed gap inductor if conductors are close to the gap (small $D_{left}$).

Then,

$$J_{eddy}(r,z,t) = -\sigma \frac{\partial A_\varphi(r,z,t)}{\partial t} \quad (20)$$

In the phasor domain, we obtain:

$$J_{eddy}(r,z,j\omega) = -j\omega\sigma A_\varphi(r,z,j\omega) \quad (21)$$

The AC resistance $R_{ac}$ is the effective resistance seen at terminals that can be obtained from dissipated power as:

$$P = R_{ac} I_L^2 = \iiint_{volume} \frac{|J_\varphi(r,\varphi,z)|^2}{\sigma} dr\, d\varphi\, dz = \frac{2\pi}{\sigma} \iint_{Area} |J_\varphi(r,z)|^2 dr\, dz \quad (22)$$

Substituting $J_\varphi$ in terms of its components leads to:

$$R_{ac} = \frac{2\pi}{\sigma} \frac{\iint_{Area} |J_s + J_{eddy}(r,z)|^2 dr\, dz}{I_L^2} \quad (23)$$

## VI. Eddy-Current Desomposition and Design Considerations

In this section, eddy current components induced in the coil due to leakage or fringing fluxes are analyzed to provide a deeper insight into the design of the inductor with minimized conductive loss. Figs. 7-10 provide flux lines, flux density and induced eddy currents at a low frequency of 1 kHz and a high frequency of 100 kHz under various scenarios. Also, Fig. 11 shows the impact of design parameters on inductors quantities including DC and AC resistances, loss components, and inductance, while there is 5 A peak as AC current for different frequencies and constant DC current of 15 A.

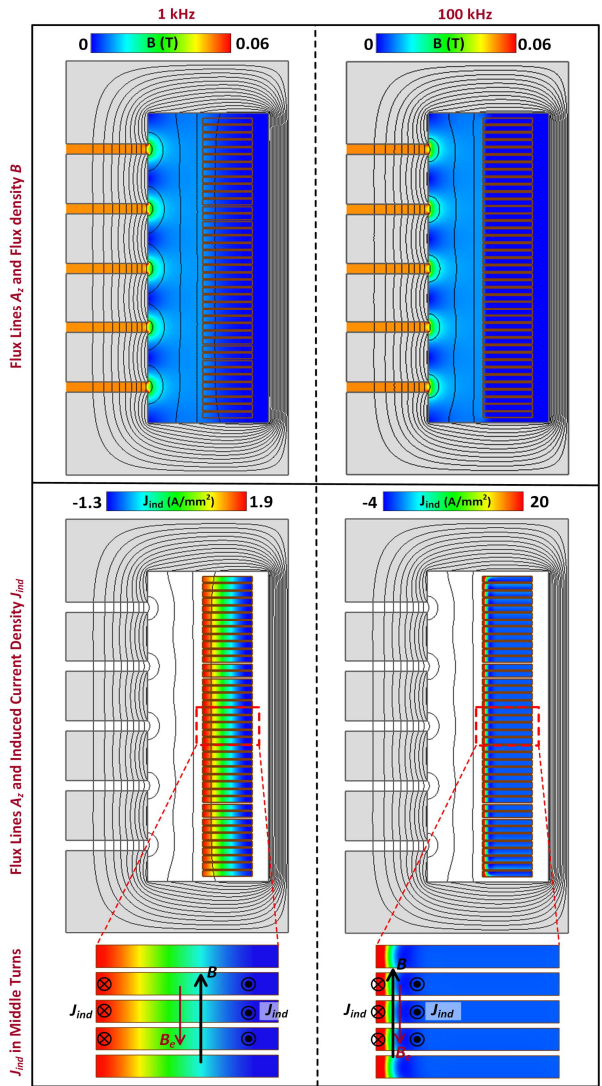

Fig. 9. Flux lines, flux density, and induced eddy currents in the conductors at frequencies of 1 kHz and 100 kHz for distributed gap inductor if conductors are away from the gap (large $D_{left}$).

As shown in Figs. 7-8, the fringing fluxes at the gap of an inductor with a single large gap flows into a large volume of conductors, while the fringing fluxes in the inductor with distributed gap are limited to smaller areas, leading to smaller eddy current losses in the conductor. It should be noted that, in order to show the closed loops of the induced eddy currents in +φ and −φ directions, the current density plots illustrate only the induced eddy currents $J_{eddy}$, not the total current $J_φ$. It can be seen that the vertical field $B$ due to gap fringing fluxes or the window leakage flux which are in the z direction induce circulating eddy currents in the r-φ plane.

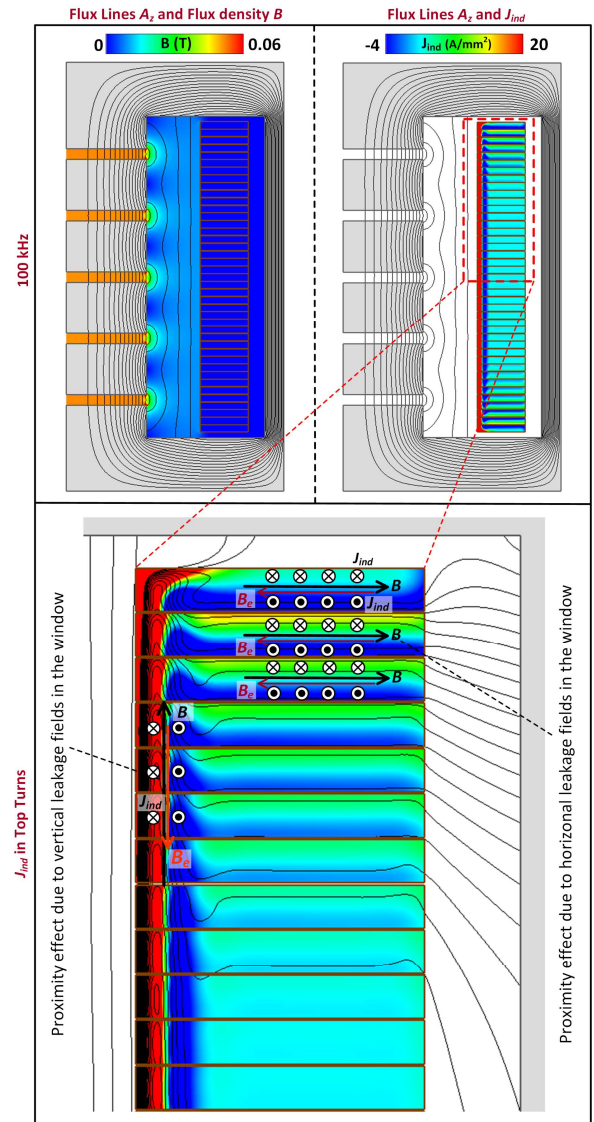

Fig. 10. Flux lines, flux density, and induced eddy currents in the conductors at frequencies of 1 kHz and 100 kHz for distributed gap inductor if conductors are very close to each other (small s).

As expected, the reaction fields $B_e$ generated by induced eddy currents are in the opposite direction of the initial field $B$ which explains how eddy currents reduce the flux linked by the coil $λ$ and thus the inductance $L = λ/i$ at higher frequencies. It can also be observed that, at 100 kHz, induced currents penetrate less into the depth of the conductor and mostly concentrate at the inner radius of the coil with a higher magnitude.

As shown in Fig. 9, there is a leakage flux within the window. The larger the total air-gap or the higher the core saturation, the larger the window leakage. In other words, if the reluctance of the main flux path within the core goes up, a larger portion of flux tends to flow into the window. These vertical fluxes cause eddy currents circulating in the r-φ plane which will be limited to the inner radius of conductors at higher frequencies, and also the window leakage flux will be confined to the inner section of the window while they also penetrate into the coil area at the lower frequencies.

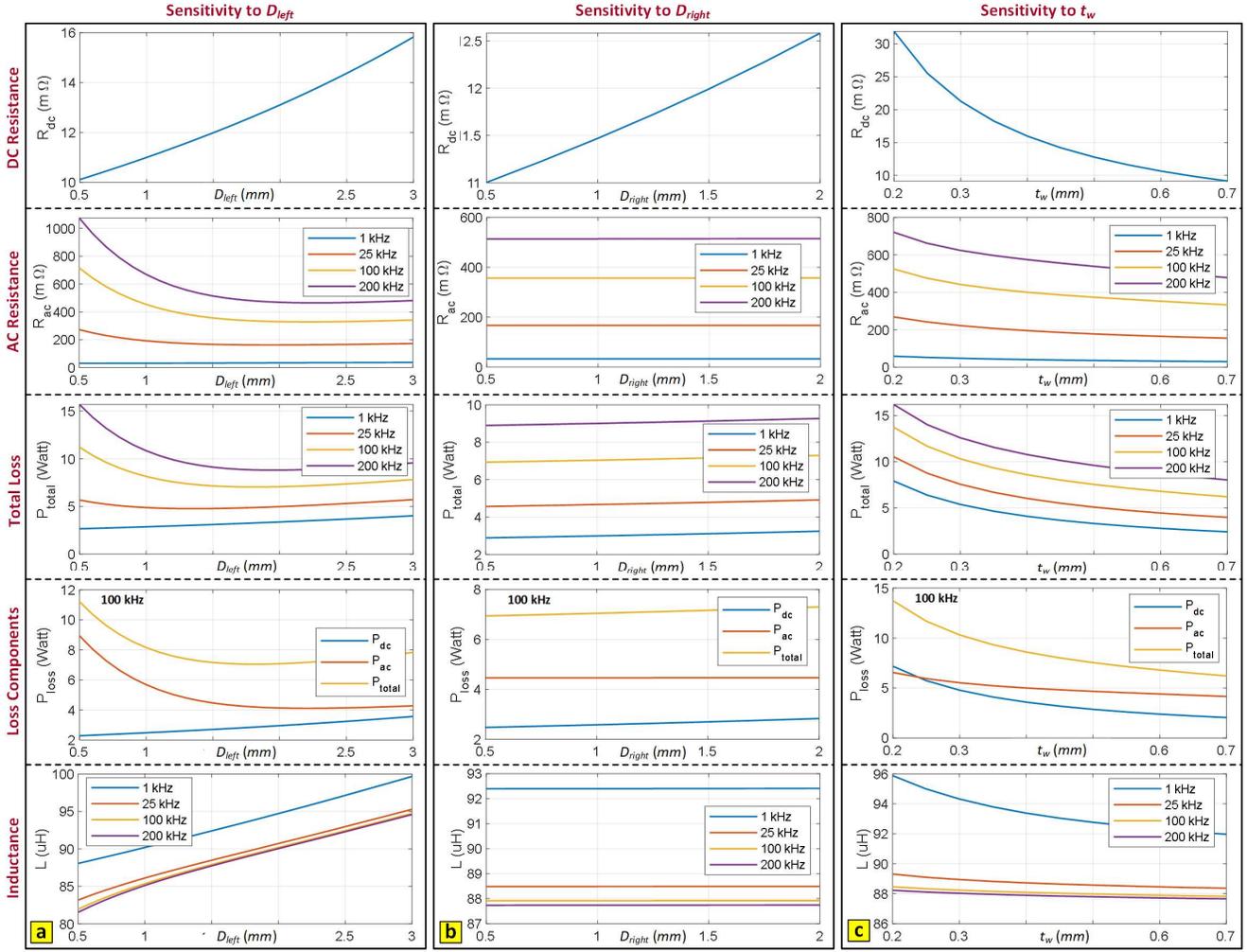

Fig. 11. Sensitivity of DC resistance, AC resistance, loss components, and inductance to design variables.

Fig. 10 represents finite element analysis for a case with small vertical distance $s$ between the turns to study proximity between coil turns. Due to the horizontal leakage fluxes in the window, eddy-currents are induced in the $z$-$\varphi$ plane.

It is seen in Fig. 11(a) that, as the coil distance from the gaps $D_{left}$ is increased, the induced eddy currents are decreased, and therefore, AC resistance and AC losses go down, while DC resistance and DC loss go up due to smaller coil cross section. It shows an optimal point to achieve the minimum total conduction loss. As leakage flux within the window is rejected by eddy currents in the conductors, this leakage flux exists only within the area between the inner radius of the coil and the center leg of the core. Therefore, the inductance linearly goes up with $D_{left}$ as this area goes up linearly. This fact shows $D_{left}$ that is a key parameter to change $L$ when N >> 1.

As shown in Fig. 11(b), the DC resistance and DC loss go up with increasing the distance between the outer radius of the coil and the core $D_{right}$. However, as the eddy currents are mostly induced on the left side of the coil near the gaps, the AC resistance and AC loss are not affected by increasing the coil distance from the right side. Additionally, the inductance is not affected by $D_{right}$ as fringing and window leakage fluxes are rejected by the inner radius of the coil. Fig. 11(c) suggests that, by increasing the coil thickness $t_w$, the DC resistance and loss goes down, and also AC resistance and AC loss go down as the skin effect goes down. However, inductance goes down due to flux rejection by the induced eddy currents.

## VII. PROTOTYPE AND EXPETRIMENTAL RESULTS

As shown in Fig. 12(a), a flat wire inductor with five distributed gaps of 1 mm each using a PQ 40/40 ferrite core is prototyped and tested, as shown in Fig. 12 (b), to verify the analytical modeling. The prototype is for an application customer spec of 87 μH ±10 % with 30 % drop at 34 A under 100 °C. The main specifications are $N = 41$, $t_w \approx 0.58$ mm, $D_w \approx 8.0$ mm, $s \approx 0.13$ mm and $r_w \approx 9.0$ mm, based on PQ 40/40 core with N95 material.

Inductance and total ESR values versus frequency are shown in Fig. 13 (a) and (b) using WYNE KERR 6500B impedance analyzer. As seen, at operating frequency of 100 kHz, $L$ is measured about 82.8 μH with 5.8 % discrepancy from value of 87.9 μH obtained by FEM. Furthermore, total ESR is measured about 500 mΩ, which here is the combination of winding's AC resistance, $R_{ac}$, discussed in previous sections, and the ESR reflected from the PQ40/40 core, $R_{core}$, due to hysteresis and eddy losses. As shown in Fig. 14(a), the lumped model of the inductor can be simplified for operating frequencies below the first resonant frequency, where $C_p$ represent the total parasitic capacitor of the inductor which models the first resonant frequency of the component, $f_r$:

$$f_r = \frac{1}{2\pi\sqrt{LC_p}} \qquad (24)$$

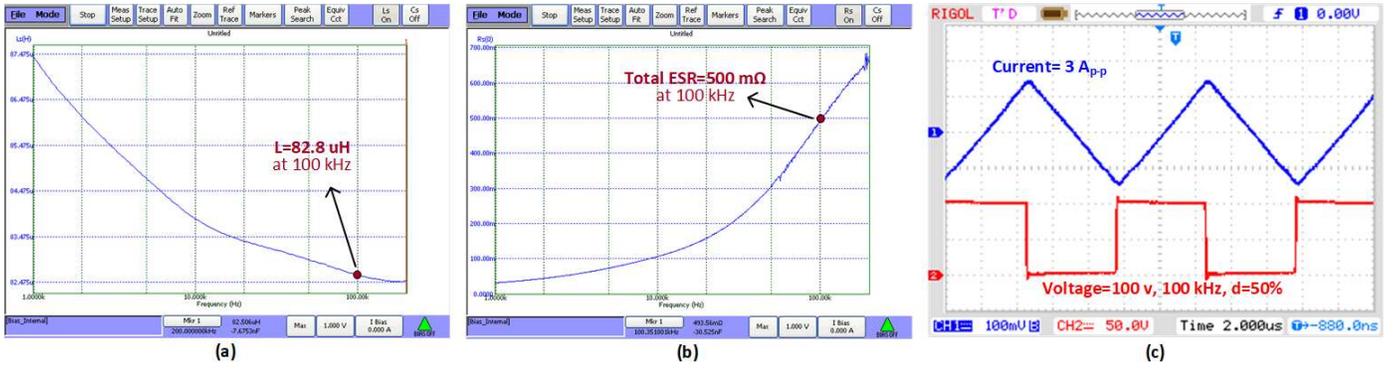

Fig. 13. Measurements of (a) inductance versus frequency, (b) AC resistance versus frequency and (c) voltage on S1 and inductor's current waveform at 100 kHz with duty cycle of =50 %.

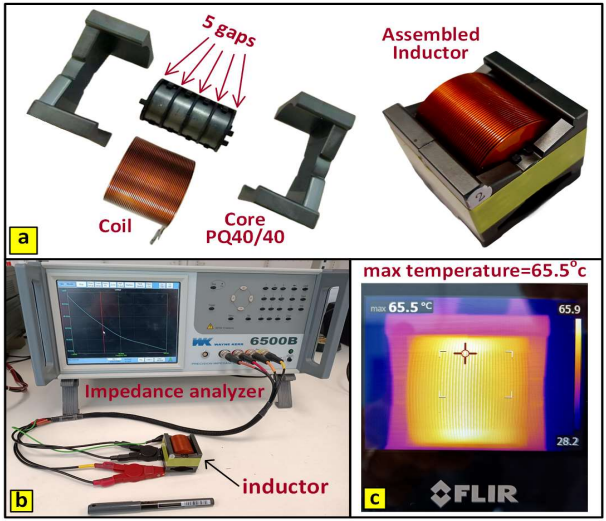

Fig. 12. Experiment: (a) prototype, (b) measurement setup, and (c) thermal camera.

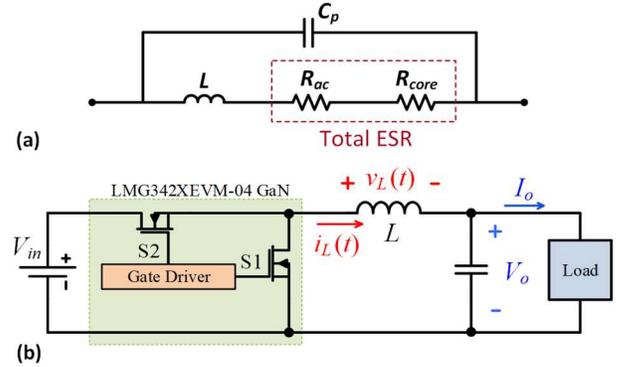

Fig. 14. Test circuit: buck converter.

To exclude $R_{core}$, first ESR of a reference inductor made by 170x0.05 mm litz wire, with $N = 41$ and the same core is measured. Regarding FEM simulations, AC and DC resistances of the litz wire are very close to each other for 100 kHz, less than 10 % deviation. Hence the measured ESR of the reference inductor is approximately equal to its DCR plus the extra ESR reflected from the core, i.e., $R_{core}$ [10].

Using WYNE KERR 6500B impedance analyser and for 100 kHz, the ESR reflected from core is estimated about 75 mΩ, by excluding DCR of the litz wire from the total ESR measured by the impedance analyser. Hence, for the flat wire prototype, ESR of the winding, $R_{ac}$, is estimated about 425 mΩ, with 19 % deviation from 357mΩ of 2D FEM results (discrepancy is due to the 3D impacts, especially large leakage fluxes in the two conductor regions without a surrounding core which causes more eddy currents losses). The DCR of the flat wire inductor is measured 12.4 mΩ, with 3.3% deviation from 12 mΩ obtained by equations (1) and (2). Due to the low DCR value, to have an accurate measurement, DCR is measured by applying DC current of 10 A and measuring the voltage drops on the inductor's terminals.

Furthermore, the inductor has been utilized in a synchronous buck converter using LMG342XEVM-04 GaN daughter card from Texas Instruments. The converter's topology is selected as a buck converter, shown in Fig. 14(b). The buck converter has been tested from 1 kHz to 200 kHz with input voltage $V_{in}$ from 10 V up to 200 V. The load is selected a variable resistive load capable of handling 0 up to 20 A.

Fig. 13 (c) shows the voltage over switch S1 and the inductor's current ripple at $f_s$ = 100 kHz, duty cycle of 50 % and input voltage of 100 V. In this condition, the current probe is connected through AC-coupling condition, showing only the ac ripple current.

Values reported for the ac inductance, included in Fig. 11, have been verified by measuring the peak-to-peak ripple current, $I_{pp}$, of the inductor and knowing the output voltage, $V_O$, and $f_s$ under duty cycle of 50 %:

$$L = \frac{V_O}{2I_{pp}f_s} \qquad (25)$$

Comparing the values derived by measurements and the values presented in Fig. 11, the maximum deviation was reported about 7.5 % for 1, 25, 100 and 200 kHz operating conditions. To illustrate the thermal performance, steady-state temperature of inductor with $f_s$ = 100 kHz, 50 % duty cycle and DC current of about 15.0 A is shown Fig. 12(c). With a temperature rise of 40 °C from the ambient temperature, 25 °C, the inductor temperature of 65 °C is still well below 100 °C which is satisfactory.

Finally, it is important to calculate the AC conduction losses of the winding for DC-DC converters. Generally, the current waveforms are triangle, as seen in Fig. 13 (c) and not a sinusoid, while the maximum ripple happens at 50 % duty cycle. Using Fourier series, voltage on the inductor $v_L(t)$ and the $h$th harmonic of the current $I_h$ are derived by:

$$v_L(t) = \sum_{h=1}^{\infty} \frac{4V_O}{\pi h} \sin(2\pi h f_s t) \qquad (26)$$

$$I_h = \frac{2V_O}{(\pi h)^2 L f_s} \qquad (27)$$

where $V_o$ is the load voltage, shown in Fig. 14 (b). Equations (26) and (27) are accurate when $hf_s < f_r$. Hence, the total ac conduction losses, $P_{ac}$, is derived by:

$$P_{ac} = \frac{1}{2} \sum_{h=1}^{\infty} R_{ac}(hf_s) I_h^2 \qquad (28)$$

where $R_{ac}(hf_s)$ is the ESR at $h$th harmonic. As discussed in [10], for the flat wire inductors, ESR is proportional to $\sqrt{f}$; hence (28) can be simplified by:

$$P_{ac} = \frac{1}{2} R_{ac} \sum_{h=1}^{\infty} \sqrt{h} I_h^2 \qquad (29)$$

By approximating (29) up to $h = 25^{th}$ harmonic, $P_{ac}$ can be simplified more in (30), where $R_{ac}$ is the winding's ESR measured at the switching frequency of the converter, $f_s$:

$$P_{ac} \approx 1.027 \left( \frac{2R_{ac}V_O^2}{\pi^4 L^2 f_s^2} \right) = 1.027 \left( \frac{8R_{ac}I_{pp}^2}{\pi^4} \right) \qquad (30)$$

Hence, $P_{ac}$ can finally be simplified based on peak-to-peak current of inductor, $I_{pp}$ and measured ESR at $f_s$.

## VIII. CONCLUSION

Analytic study and design considerations for flat wire inductors are presented. Magnetic equivalent circuits including frequency-dependent reluctances and inductance are introduced. Three formulations for DCR are derived which can be used depending on the coil topology and the required accuracy. Calculations for ESR and AC losses are presented. Effectiveness of distributed gaps in terms of reduced fringing flux and thus eddy current losses is shown. Eddy current components are scrutinized. Sensitivity of inductor quantities such as DCR, ESR, loss components and inductance to design parameters are studied. The optimum design and the balance between DCR and ESR are discussed. Finally, an inductor is prototyped, where the experimental results are on par with FEM and show the satisfactory performance of the design.

There is a growing need for interpretable, generalizable and low-cost AI models, while physics-based models face their own challenges. In a future work, such physics-based analytical models can be integrated into the architecture, loss function, or optimization constraints of AI/ML models as physics-informed artificial intelligence (PIAI) to leverage both data-driven and physics-based techniques.